\documentclass[hidelinks,conference]{ieeeconf}      

\IEEEoverridecommandlockouts                              

\usepackage{xr}

\newcommand{\omitted}[1]{}

\title{
\fontsize{18}{18} \selectfont 
Safe Non-Stochastic Control of Linear Dynamical Systems
}

\author{Hongyu Zhou, Vasileios Tzoumas
	\thanks{Department of Aerospace Engineering, University of Michigan, Ann Arbor, MI 48109 USA;  {\tt\footnotesize \{zhouhy, vtzoumas\}@umich.edu}}
	\vspace{-8mm}
}

\usepackage{cite}

\usepackage{comment}
\usepackage{siunitx}
\usepackage{relsize}
\usepackage{ifthen}
\usepackage[colorinlistoftodos]{todonotes}

\usepackage[caption=false]{subfig}

\usepackage[vlined,ruled,linesnumbered]{algorithm2e}
\usepackage{graphics} 
\usepackage{rotating}
\usepackage{color}
\usepackage{enumerate}
\usepackage[T1]{fontenc}
\usepackage{psfrag}
\usepackage{epsfig} 
\usepackage{booktabs}
\usepackage{graphicx,url}
\usepackage{multirow}
\usepackage{array}
\usepackage{latexsym}
\usepackage{amsfonts}
\usepackage{amsmath}
\usepackage{amssymb}
\usepackage{xstring}
\usepackage{algorithmic}
\usepackage{multirow}
\usepackage{xcolor}
\usepackage{prettyref}
\usepackage{flexisym}
\usepackage{bigdelim}
\usepackage{breqn} 
\usepackage{listings}
\let\labelindent\relax
\usepackage{xspace}
\usepackage{bm}
\graphicspath{{./figures/}}
\usepackage{tikz}
\usetikzlibrary{matrix,calc}
\usepackage{lipsum}
\usepackage{mdwlist}

\let\itemize\undefined
\makecompactlist{itemize}{stditemize}
\usepackage{enumitem}
\usepackage{caption}
\usepackage{amsthm}
\usepackage{mathtools}

\makeatletter
\let\NAT@parse\undefined
\makeatother
\usepackage{hyperref}
\usepackage{cleveref}

\hypersetup{%
  colorlinks=true,
  linkcolor=blue,
  filecolor=magenta,      
  urlcolor=blue,
  citecolor=red,
  linkbordercolor={0 0 1}
}

\newtheorem{theorem}{Theorem}
\newtheorem{problem}{Problem}

\newtheorem{lemma}{Lemma}
\newtheorem{assumption}{Assumption}
\newtheorem{definition}{Definition}

\newtheorem{remark}{Remark}

\newcommand{\bdmath}{\begin{dmath}}
\newcommand{\edmath}{\end{dmath}}
\newcommand{\beq}{\begin{equation}}
\newcommand{\eeq}{\end{equation}}
\newcommand{\bdm}{\begin{displaymath}}
\newcommand{\edm}{\end{displaymath}}
\newcommand{\bea}{\begin{eqnarray}}
\newcommand{\eea}{\end{eqnarray}}
\newcommand{\beal}{\beq \begin{array}{lll}}
\newcommand{\eeal}{\end{array} \eeq}
\newcommand{\beas}{\begin{eqnarray*}}
\newcommand{\eeas}{\end{eqnarray*}}
\newcommand{\ba}{\begin{array}}
\newcommand{\ea}{\end{array}}
\newcommand{\bit}{\begin{itemize}}
\newcommand{\eit}{\end{itemize}}
\newcommand{\ben}{\begin{enumerate}}
\newcommand{\een}{\end{enumerate}}

\newcommand{\calK}{{\cal K}}

\newcommand{\calO}{{\cal O}}

\newcommand{\calS}{{\cal S}}

\newcommand{\calU}{{\cal U}}

\newcommand{\calW}{{\cal W}}
\newcommand{\calX}{{\cal X}}
\newcommand{\calZ}{{\cal Z}}

\definecolor{myblue}{RGB}{65 105 225}

\newcommand{\hide}[1]{}

\newcommand{\hiddenText}{{\color{gray} hidden text.}}
\newcommand{\hideWithText}[1]{\hiddenText}

\newcommand{\vs}{\boldsymbol{s}}

\newcommand{\scenario}[1]{{\fontsize{9}{8.7}\selectfont\sf#1}\xspace}

\newcommand{\TimeSeq}{\{1, \dots, T\}}

\newcommand{\myIdentity}{\mathbf{I}}

\newcommand{\myx}{\mathbf{x}}
\newcommand{\myv}{\mathbf{v}}
\newcommand{\SumOneT}{\sum_{t=1}^{T}}

\renewcommand{\vs}{\emph{vs.}\xspace}
\newcommand{\ie}{\emph{i.e.},\xspace}
\newcommand{\eg}{\emph{e.g.},\xspace}

\newcommand{\red}[1]{{\color{red}#1}}
\newcommand{\blue}[1]{{\color{blue}#1}}

\newcommand{\myParagraph}[1]{{\bf #1.}\xspace}

\newcommand{\OGD}{\scenario{\textbf{OGD}}}

\newcommand{\OCO}{\scenario{{OCO}}}
\newcommand{\OCOM}{\scenario{{OCO-M}}}
\newcommand{\OCOTV}{\scenario{{OCO-TV}}}
\newcommand{\SafeOCO}{\scenario{{Safe-OCO}}}
\newcommand{\SafeNSC}{\scenario{{Safe-NSC}}}
\newcommand{\SafeOGD}{\scenario{\textbf{Safe-OGD}}}

\newcommand{\DAC}{\scenario{\textbf{DAC}}}

\newcommand{\DyReg}{\operatorname{Regret}_T^D}

\newcommand{\DyRegNSC}{\operatorname{Regret-NSC}_T^D}

\renewcommand{\algorithmicrequire}{\textbf{Input:}}
\renewcommand{\algorithmicensure}{\textbf{Output:}}

\PassOptionsToPackage{end}{algorithmic}

\renewcommand{\baselinestretch}{0.96}

\begin{document}








\maketitle

\thispagestyle{empty}
\pagestyle{empty}

\begin{abstract}
We study the problem of \textit{safe control of linear dynamical systems corrupted with non-stochastic noise}, and provide an algorithm that guarantees (i) zero constraint violation of convex time-varying constraints,  and (ii) bounded dynamic regret, \ie bounded suboptimality against an optimal clairvoyant controller that knows the future noise a priori. 
The constraints bound the values of the state and of the control input such as to ensure collision avoidance and bounded control effort.
We are motivated by the future of autonomy where robots will safely perform complex tasks despite real-world  unpredictable disturbances such as wind and wake disturbances.
To develop the algorithm, we capture our problem as a sequential game between a linear feedback controller and an adversary, assuming a known upper bound on the noise's magnitude.
Particularly, at each step $t=1,\ldots, T$, first the controller chooses a linear feedback control gain $K_t \in \calK_t$, where $\calK_t$ is constructed such that it guarantees that the safety constraints will be satisfied;
then, the adversary reveals the current noise $w_t$ and the controller suffers a loss $f_t(K_t)$ ---\eg $f_t$ represents the system's tracking error at $t$ upon the realization of the noise. The controller aims to minimize its cumulative loss, despite knowing $w_t$ only after $K_t$ has been chosen.
We validate our algorithm in simulated scenarios of safe control of linear dynamical systems in the presence of bounded noise. 
\end{abstract}

\begin{tikzpicture}[overlay, remember picture]
\path (current page.north east) ++(-1.6,-0.2) node[below left] {
	A preliminary version of this paper has been accepted for publication in the IEEE Conference on Decision and Control (CDC).
};
\end{tikzpicture}

\begin{tikzpicture}[overlay, remember picture]
\path (current page.north east) ++(-2.4,-0.6) node[below left] {
		 The CDC paper does not contain Appendices A, C, and D, and the proofs of Lemma 1, Lemma 2, and Theorem 1.
};
\end{tikzpicture}

\begin{tikzpicture}[overlay, remember picture]
\path (current page.north east) ++(-2.8,-1.0) node[below left] {
Please cite the paper as:
	H.~Zhou, V.~Tzoumas, ``Safe Non-Stochastic Control of Linear Dynamical Systems”,
};
\end{tikzpicture}
\begin{tikzpicture}[overlay, remember picture]
\path (current page.north east) ++(-6.5,-1.4) node[below left] {
		 IEEE Conference on Decision and Control (CDC), 2023.
};
\end{tikzpicture}

\section{Introduction}\label{sec:Intro}

In the future, robots will be leveraging their on-board control capabilities to complete safety-critical tasks such as package delivery~\cite{ackerman2013amazon}, target tracking~\cite{chen2016tracking}, and disaster response~\cite{rivera2016post}.
To complete such complex tasks, the robots need to \textit{reliably} overcome a series of key challenges:

\vspace{.5mm}
\paragraph{Challenge I: Time-Varying Safety Constraints} The robots need to ensure their own safety and the safety of their surroundings.  For example,  robots often need to ensure that they follow prescribed collision-free trajectories or that their control effort is kept under prescribed levels. Such safety requirements take the form of \textit{time-varying} state and control input constraints, and can make the {planning of control inputs computationally hard~\cite{rawlings2017model,borrelli2017predictive}.}

\paragraph{Challenge II: Unpredictable Noise} The robots' dynamics are often corrupted by unknown and non-stochastic noise, \ie noise that is not necessarily i.i.d.~Gaussian and, broadly, stochastic.  For example, aerial and marine vehicles often face non-stochastic winds and waves, respectively~\cite{faltinsen1993sea, sapsis2021statistics}.  But the current control algorithms primarily rely on stochastic noise (\eg Gaussian-structured), compromising thus the robots' ability to ensure safety~\cite{aastrom2012introduction,berkenkamp2019safe}.
\vspace{.5mm}
    

The above challenges motivate the development of safe control algorithms against unpredictable noise. State-of-the-art methods that aim to address this problem either rely on robust control~\cite{goel2020regret,sabag2021regret,goel2021regret,martin2022safe,didier2022system,zhou2022safe} or on online learning for control~\cite{hazan2020nonstochastic,agarwal2019online,li2021online,simchowitz2020improper,gradu2020adaptive,zhao2021non,zhou2023efficient}. But the robust methods are often conservative and computationally heavy since they simulate the system dynamics over a lookahead horizon assuming a worst-case noise realization that matches a known upper bound on the magnitude of the noise.
To reduce conservatism and increase efficiency, researchers have recently focused on online learning for control methods via Online Convex Optimization (\OCO) \cite{shalev2012online,hazan2016introduction}. The online learning for control methods typically rely on the \textit{Online Gradient Descent} (\OGD) algorithm and its variants, offering \textit{bounded regret} guarantees, \ie bounded suboptimality with respect to an optimal (possibly time-varying) clairvoyant controller~\cite{hazan2020nonstochastic,agarwal2019online,li2021online,simchowitz2020improper,gradu2020adaptive,zhao2021non,zhou2023efficient}. However, the current online methods address only time-\underline{in}variant safety constraints.

\myParagraph{Contributions} Our goal is to achieve online control of linear dynamical systems subject to time-varying safety constraints, despite unpredictable noise.
To this end, we formalize the problem of \textit{Safe Non-Stochastic Control of Linear Dynamical Systems} (\SafeNSC). \SafeNSC can be interpreted as a sequential game between a linear feedback controller and an adversary. Particularly, at each step $t=1,\ldots, T$, first the controller chooses a linear feedback control gain $K_t \in \calK_t$, where $\calK_t$ is constructed such that it guarantees that the safety constraints will be satisfied;
then, the adversary reveals the current noise $w_t$ and the controller suffers a loss $f_t(K_t)$ ---\eg $f_t$ represents the system's tracking error at $t$ upon the realization of the noise. 

\SafeNSC is challenging since the controller aims to guarantee bounded dynamic regret despite knowing $w_t$ only after $K_t$ has been chosen.

We make the following contributions to solving \SafeNSC:

\begin{itemize}

     \item \textit{Algorithmic Contributions}: 
     We introduce the algorithm \textit{Safe Online Gradient Descent} (\SafeOGD), which generalizes the seminal \textit{Online Gradient Descent} (\OGD) \cite{zinkevich2003online} to the \SafeNSC setting to enable online non-stohastic control subject to time-varying constraints.

     \item \textit{Technical Contributions}: We prove that the \SafeOGD controller has bounded dynamic regret against any safe linear feedback control policy (\Cref{theorem:SafeOGD_Control}), given a known upper bound on the noise's magnitude. 
     When the domain sets are time-\underline{in}variant, we prove that the bound of \SafeOGD reduces to the bound in the standard (time-\underline{in}variant) \OCO setting \cite{zinkevich2003online} (\Cref{subsec:invariant}).
     
\end{itemize}

\myParagraph{Numerical Evaluations} {
We validate our algorithms in simulated scenarios.
(\Cref{sec:exp} and \Cref{app:supp_exp}).  
Specifically, we compare our algorithm with the safe $H_2$ and $H_\infty$ in scenarios involving a quadrotor aiming to stay at a hovering position despite unpredictable disturbances (\Cref{sec:exp}). We then compare our algorithm with state-of-the-art \OCO with Memory (\OCOM) algorithm~\cite{agarwal2019online} in scenarios involving synthetic
linear time-invariant systems (\Cref{app:supp_exp}).
{
Our algorithm achieves in the simulations (i) comparable loss and better computation time performance than safe $H_2$ and $H_\infty$, and (ii) better loss performance than \OCOM algorithm.
}
\section{Related Work}\label{sec:lit_review}

We next review (i) \textit{Non-Stochastic Control}: regret optimal control and online learning for control; (ii) \textit{\OCO} with \textit{time-\underline{in}variant constraints} or \textit{time-varying constraints}.

\myParagraph{{Regret-optimal} control}
Regret optimal control algorithms select control inputs upon simulating the future system dynamics across a {lookahead} horizon~\cite{goel2020regret,sabag2021regret,goel2021regret,martin2022safe,didier2022system,zhou2022safe}. Specifically, these methods guarantee safety via solving a robust optimization problem that (ii-a) assumes a worst-case realization of noise, which can be pessimistic and time-consuming, thus compromising the quality of real-time control; and (ii-b) requires the a priori knowledge of a closed-form solution of the optimal controller over the lookahead horizon, which is not available in the safe non-stochastic control setting.

\myParagraph{Online learning for control} Online learning algorithms select control inputs based on past information only~\cite{hazan2020nonstochastic,agarwal2019online,li2021online,simchowitz2020improper,gradu2020adaptive,zhao2021non,zhou2023efficient}. By employing the \OCO framework, they provide bounded regret guarantees against an optimal (potentially time-varying) clairvoyant controller even though the noise is unpredictable. However, they consider time-\underline{in}variant state and control input constraints, in contrast to time-varying safety constraints in this paper.  In more detail: 

\setcounter{paragraph}{0}

\paragraph{OCO with Time-\underline{In}variant Constraints} We focus our review on algorithms with bounded dynamic regret; for a broader review of \OCO algorithms, we refer the reader to~\cite{hazan2016introduction}. \cite{zhang2018adaptive} prove that the optimal dynamic regret for \OCO is $\Omega\left(\sqrt{T(1+C_{T})}\right)$, {where $C_T \triangleq \sum_{t=2}^{T} \| \myv_{t-1} - \myv_{t} \|$ is the path length of the comparator sequence}, and provide an algorithm matching this bound.  The algorithm is based on \textit{Online Gradient Descent} (\OGD)~\cite{zinkevich2003online}, which is a projection-based algorithm: at each time step $t$, \OGD chooses a decision $\myx_{t}$ by first computing an intermediate decision $\mathbf{x}_{t}^\prime=\myx_{t-1}-\eta\nabla f_{t-1}(\myx_{t-1})$ ---given the previous decision $\myx_{t-1}$, the gradient of the previously revealed loss $f_{t-1}(\myx_{t-1})$, and a step size $\eta>0$--- and then projects $\mathbf{x}_{t}^\prime$ back to the time-\underline{in}variant domain set $\calX$ to output the final decision $\mathbf{x}_{t}$. 

\paragraph{OCO with Time-Varying Constraints}
The problem of \textit{\OCO with Time-Varying Constraints} (\OCOTV) is defined as follows: 
At each time step $t$, first the optimizer chooses a decision $\myx_t$ from a time-\underline{in}variant domain set $\calX$, and then the adversary reveals a convex loss function $f_t$ as well as a vector-valued constraint  $g_t(\myx_t)\leq 0$. The optimizer in particular aims to minimize (i) the cumulative loss $\sum_{t=1}^{T} f_t(\myx_t)$, and (ii) the cumulative constraint violation $\sum_{t=1}^{T} g_t(\myx_t)$. In contrast thus to the setting in this paper, where the constraints must be satisfied at each time step, in the \OCOTV setting the optimizer may violate any of the constraints, aiming to only asymptotically guarantee in the best case no-regret constraint violation, \ie $\lim_{T\rightarrow\infty}\sum_{t=1}^{T} g_t(\myx_t)/T=0$ \cite{cao2018online,chen2018bandit,yi2020distributed,mahdavi2012trading,carnevale2022distributed,paternain2016online}.

\section{Problem Formulation}\label{sec:problem}

We formulate the problem of \textit{Safe Non-Stochastic Control of Linear Dynamical Systems} (\Cref{prob:SafeNSC}). To this end, we use the following notation and assumptions.

\myParagraph{Linear Time-Varying System}
We consider Linear Time-Varying (LTV) systems of the form
\begin{equation}
	x_{t+1} = A_{t} x_{t} + B_{t} u_{t} + w_{t}, \quad t=1, \ldots, T,
	\label{eq:LTV}
\end{equation}
where $x_t \in\mathbb{R}^{d_x}$ is the state of the system, $u_t \in\mathbb{R}^{d_u}$ is the control input, and $w_t\in\mathbb{R}^{d_x}$ is the process noise.  
\begin{assumption}[Known System Matrices]
    The system matrices, \ie $A_t$ and $B_t$, are known.
\end{assumption}

\begin{assumption}[Bounded System Matrices and Noise]\label{assumption:bounded_system_noise}
The system matrices and noise are bounded, \ie $\|A_t\|\; \leq \kappa_{A}$, $\|B_t\|\; \leq \kappa_{B}$, and $w_t \in \calW \triangleq \{w \mid \left\|w\right\| \leq W\}$, where $\kappa_{A}$, $\kappa_{B}$, and $W$ are given positive numbers.
\end{assumption}

Per \Cref{assumption:bounded_system_noise}, we assume no stochastic model for the process noise $w_t$.  The noise may even be adversarial, subject to the bounds prescribed by $W$.



\myParagraph{Safety Constraints} We consider the states and control inputs for all $t$ must satisfy polytopic constraints of the form
\begin{equation}
    \begin{aligned}
        & x_t \in \calS_t \triangleq \{x \mid L_{x,t} x \leq l_{x,t}\}, \ \forall \{{w}_{\tau} \in \calW\}_{\tau=1}^{t-1}, \\
        & u_t \in \calU_t \triangleq \{u \mid L_{u,t} u \leq l_{u,t}\}, 
    \end{aligned}
    \label{eq:safety_constraints}
\end{equation}
for given $L_{x,t}$, $l_{x,t}$, $L_{u,t}$, and $l_{u,t}$.\footnote{Our results hold true also for any convex state and control input constraints. We focus on polytopic constraints for simplicity in the presentation.}

\myParagraph{Linear-Feedback Control Policy} We consider a linear state feedback control policy $u_{t} = K_{t} x_{t}$ such that
\begin{equation}
     \| K_t \|\; \leq \kappa, \   \| A_t - B_t K_t\|\; \leq 1-\gamma,
    \label{eq:Kt}
\end{equation}
for given $\kappa > 0$
and $\gamma \in (0,1)$, 
where $K_t$ will be optimized online. The constraint $\| K_t \|\; \leq \kappa$ ensures $K_t$ is chosen from a compact decision set
, and the constraint $\| A_t - B_t K_t\|\; \leq 1-\gamma$ ensures the state is bounded for all $t$; both constraints
enable bounding the dynamic regret of the proposed online optimization algorithm.  To ensure that also the safety constraints in \cref{eq:safety_constraints} are satisfied, we impose additional constraints on $K_t$ later in the paper (\Cref{lemma:calK_t} presented in \Cref{sec:SafeOCO-Alg}).

\begin{remark}[Removal of the constraint $\| A_t - B_t K_t\|\; \leq 1-\gamma$]
The constraint can be removed by employing a sequence $K_t^s$ of $(\epsilon, \gamma)$ sequentially stabilizing controllers, \ie setting $u_t = - K_t x_t -K_t^s x_t$, where $K_t^s$ is sequentially stabilizing~\cite{gradu2020adaptive} and $\| K_t x_t \|$ is bounded.
\end{remark}

\myParagraph{Loss Function} We consider loss functions (control costs) that satisfy the following assumption.

\begin{assumption}[Convex and Bounded Loss Function with Bounded Gradient]\label{assumption:cost}
$c_{t}(x_{t+1},u_t): \mathbb{R}^{d_{x}} \times \mathbb{R}^{d_{u}} \rightarrow \mathbb{R}$ is convex in $x_{t+1}$ and $u_t$. Further, when $\|x\|\,\leq D$, $\|u\|\leq D$ for some $D>0$, then $\left|c_{t}(x, u)\right| \leq\beta D^{2}$ and $\left\|\nabla_{x} c_{t}(x, u)\right\|\leq G D,\left\|\nabla_{u} c_{t}(x, u)\right\| \leq G D$, for given  $\beta$ and $G$.
\end{assumption} 

{An example of a loss function that satisfies \Cref{assumption:cost} is the quadratic loss $c_{t}\left(x_{t+1}, u_{t}\right) = x_{t+1} Q x_{t+1}^\top + u_{t} R u_{t}^\top$.}

\myParagraph{Control Performance Metric} We design the control inputs $u_t$ to ensure both safety and a control performance comparable to an optimal clairvoyant policy that selects $u_t$ knowing the future noise realizations ${w}_t$ a priori. 

\begin{definition}[Dynamic Policy Regret]\label{def:DyReg_control}
The \emph{dynamic policy regret} is defined as follows:
\begin{equation}
	\DyRegNSC = \sum_{t=1}^{T} c_{t}\left(x_{t+1}, u_{t}\right)-\sum_{t=1}^{T} c_{t}\left(x_{t+1}^{*}, u_{t}^{*}\right),
	\label{eq:DyReg_control}
\end{equation}
{where (i) both sums in \cref{eq:DyReg_control} are evaluated with the same noise $\{w_1,\ldots, w_T\}$,  which is the noise experienced by the system during its evolution per the control inputs $\{u_1,\ldots, u_T\}$, (ii) $u_{t}^{*} \triangleq -K_t^{*} x_{t}$ is the optimal linear feedback control input in hindsight, {\ie the optimal input given a priori knowledge of $w_t$}, (iii) $x_{t+1}^{*} \triangleq A_t x_t + B_t u_{t}^{*} + w_t$ is the state reached by applying the optimal control inputs $u_{t}^{*}$ from state $x_{t}$, and (iv) $x_{t+1}^{*}$ and $u_{t}^{*}$ satisfy constraints in \cref{eq:safety_constraints} for all $t$.}
\end{definition}

\myParagraph{Problem Definition} We formally define the problem of \textit{Safe Non-Stochastic Control of Linear Dynamical Systems}:
\begin{problem}[Safe Non-Stochastic Control of Linear Dynamical Systems (\SafeNSC)]\label{prob:SafeNSC}
Assume the initial state of the system is safe, \ie $x_0 \in \calS_{0}$. At each $t=1,\ldots, T$, first a control input $u_t \in \calU_t$ is chosen; then, {the noise $w_t \in \mathbb{R}^{d_{x}}$ is revealed}, the system evolves to state $x_{t+1} \in \calS_{t+1}$, and suffers a loss $c_t(x_{t+1},u_t)$. The goal is to guarantee states and control inputs that satisfy the constraints in \cref{eq:safety_constraints} for all $t$ and that minimize the dynamic policy regret.
\end{problem}

\section{\SafeOGD Algorithm} \label{sec:SafeOCO-Alg}

\begin{algorithm}[t]
	\caption{\mbox{\hspace{-.05mm}Safe Online Gradient Descent (\SafeOGD)} for \SafeNSC (\Cref{prob:SafeNSC}).}
	\begin{algorithmic}[1]
        \REQUIRE Time horizon $T$; step size $\eta$.
        \ENSURE Control ${u}_t$ at each time step $t=1,\ldots,T$.
        \medskip
        \STATE Initialize ${K}_1 \in \calK_1$; 
        \FOR {each time step $t = 1, \dots, T$}
        \STATE Output $u_t = -K_t x_t$;
        \STATE Observe the state $x_{t+1}$ and calculate the noise $w_t = x_{t+1} - A_t x_t - B_t u_t$;
        \STATE Suffer the loss $c_t(x_{t+1}, u_t)$;
        \STATE Express the loss function in $K_t$ as $f_t({K}_t): \mathbb{R}^{d_u\times d_x} \rightarrow \mathbb{R}$;
        \STATE Obtain gradient $\nabla_K f_t({K}_t)$;
        \STATE Obtain domain set $\calK_{t+1}$;
        \STATE Update $K_{t+1}^\prime=K_t- \eta \nabla_K f_t(K_t)$;
        \STATE Project $K_{t+1} = \Pi_{\calK_{t+1}}(K_{t+1}^\prime)$;
		\ENDFOR
	\end{algorithmic}\label{alg:SafeOGD_Control}
\end{algorithm}

We present \SafeOGD (\Cref{alg:SafeOGD_Control}) with bounded dynamic regret for \SafeNSC.
\SafeOGD first initializes $K_1 \in \calK_1$, where $\calK_1$ is defined per \Cref{lemma:calK_t} (line 1). At each iteration $t$, \Cref{alg:SafeOGD_Control} evolves to state $x_{t+1}$ with the control inputs $u_t = -K_t x_t$ and obtain the noise $w_t$ (lines 3-4). After that, the cost function is revealed and the algorithm suffers a loss of $c_t(x_{t+1},u_t)$ (line 5). Then, \SafeOGD expresses $c_t(x_{t+1},u_t)$ as a function of $K_t$, denoted as $f_t(K_t) \triangleq c_t((A_t - B_t K_t) x_t + w_t, - K_t x_t)$ ---which is convex in $K_t$, given $A_t$, $B_t$, $x_t$, and $w_t$, per \Cref{lemma:cvx_K} below--- and obtains the gradient $\nabla_K f_t(K_t)$ (lines 6-7). To ensure safety, \SafeOGD constructs the domain set $\calK_{t+1}$ per \Cref{lemma:calK_t} (line 8), {which requires one step ahead knowledge of $L_{x,t+1}, L_{u,t+1}, l_{x,t+1}, l_{u,t+1}$}. Finally, \SafeOGD updates the control gain and projects it back to $\calK_{t+1}$ (lines 9-10).

\begin{lemma}[Convexity of Loss function in Control Gain]\label{lemma:cvx_K}
The loss function $c_{t}\left(x_{t+1}, u_{t}\right) : \mathbb{R}^{d_x} \times \mathbb{R}^{d_u} \rightarrow \mathbb{R}$ is convex in $K_t$.
\end{lemma}
\noindent\textit{Proof:} The proof follows by the convexity of $c_{t}(x_{t+1},u_t): \mathbb{R}^{d_{x}} \times \mathbb{R}^{d_{u}} \rightarrow \mathbb{R}$ in $x_{t+1}$ and $u_t$, and the linearity of $x_{t+1}$ and $u_t$ in $K_t$, \ie $x_{t+1} = A_t x_t + B_t u_t + w_t$ and $u_t = -K_t x_t$ given $A_t$, $B_t$, $x_t$ , and $w_t$.
\qed

\begin{lemma}[Set of Control Gains that Guarantee Safety]\label{lemma:calK_t}
By choosing $K_t \in \calK_t$, where
\begin{equation}
    \begin{aligned}
    \calK_t &\triangleq \{  K \mid - L_{x,t} B_t K x_t  \leq l_{x,t} - L_{x,t} A_t x_t - W\|L_{x,t}\|, \\
     & -L_{u,t} K x_t \leq l_{u,t}, \  \| K \|\; \leq \kappa, \   \| A_t - B_t K\| \;\leq 1-\gamma \},
    \end{aligned}
    \label{eq:lemma_calK_t}
\end{equation}
then, $x_{t+1} \in \calS_{t+1}$ and $u_t \in \calU_t$ at each time step $t$.
\end{lemma}

\noindent\textit{Proof:} At time step $t$, we aim to choose $K_t$ such that the safety constraints on state $x_{t+1}$ and control input $u_t$ are satisfied, \ie
\begin{equation}
    \begin{aligned}
        x_{t+1} &= A_t x_t + B_t u_t + w_t \\
        &\in \calS_{t+1} \triangleq \{x \mid L_{x,t+1} x \leq l_{x,t+1}\}, \ \forall {w}_{t} \in \calW, \\
        u_t &\in \calU_t \triangleq \{u \mid L_{u,t} u_t \leq l_{u,t}\}, 
    \end{aligned}
    \label{eq:lemma_calK_t_1}
\end{equation}
for given $A_t$, $B_t$, $x_t$, $L_{x,t+1}$, $l_{x,t+1}$, $L_{u,t}$, $l_{u,t}$, and control input $u_t = -K_t x_t$.
Hence, \cref{eq:lemma_calK_t_1} can be rewritten as
\begin{equation}
    \begin{aligned}
        L_{x,t+1} A_t x_t - L_{x,t+1} B_t K_t x_t + L_{x,t+1} w_t  &\leq l_{x,t+1}, \ \forall {w}_{t} \in \calW, \\
        -L_{u,t} K_t x_t &\leq l_{u,t}.
    \end{aligned}
    \label{eq:lemma_calK_t_2}
\end{equation}
By applying now robust optimization \cite{ben2009robust}, \cref{eq:lemma_calK_t_2} becomes
\begin{equation}
    \begin{aligned}
        - L_{x,t+1} B_t K_t x_t  &\leq l_{x,t+1} - L_{x,t+1} A_t x_t - W\|L_{x,t+1}\|, \\
        -L_{u,t} K_t x_t &\leq l_{u,t}. 
    \end{aligned}
    \label{eq:lemma_calK_t_2}
\end{equation}
Combining \cref{eq:Kt,eq:lemma_calK_t_2}, we construct the domain set $\calK_t$ as in \cref{eq:lemma_calK_t}, which is also
 convex in $K_t$.
\qed

\begin{assumption}[Recursive Feasibility]\label{assump:calK_t}
We assume that the domain set $\calK_t$ is non-empty for all $t$, $t \in \TimeSeq$.\footnote{The discussion on recursive feasibility is given in \Cref{app:rf}.}
\end{assumption}

\section{Dynamic Regret Analysis}\label{sec:SafeOCO-Reg}

We present the dynamic regret bound for \SafeOGD against any comparator sequence (\Cref{theorem:SafeOGD_Control}).  The bound reduces to the bound of standard OCO when the optimization domain is time-\underline{in}variant (\Cref{remark:bound_reduction} in \Cref{subsec:invariant}).
We use the notation:
\begin{itemize}
{
    \item $\Pi_{\calK}(\cdot)$ is a projection operation onto the set $\calK$;
}
    \item $\bar{K}_{t+1} \triangleq \Pi_{\calK_{t}}(K_{t+1}^\prime)$ is the decision would have been chosen at time step $t+1$ if $\calK_{t+1} = \calK_{t}$;

    \item $\zeta_t \triangleq \left\|\bar{K}_{t+1} - {K}_{t+1} \right\|_{\mathrm{F}}$ 
    is the distance between $\bar{K}_{t+1}$ and ${K}_{t+1}$, which are the projection of ${K}_{t+1}^\prime$ onto sets $\calK_t$ and $\calK_{t+1}$, respectively.  Thus, it quantifies how fast the safe domain set changes ---$\zeta_t$ is  $0$ when $\calK_t=\calK_{t+1}$;

    \item $S_T \triangleq \SumOneT \zeta_t$ is the cumulative variation of decisions due to time-varying domain sets ---$S_T$ becomes $0$ when domain sets are time-\underline{in}variant; 

    \item $C_T \triangleq \sum_{t=2}^{T} \| K_{t-1}^\star - K_{t}^\star \|_{\mathrm{F}}$ is the path length of the sequence of comparators.  It quantifies how fast the optimal control gains change.
\end{itemize}

\subsection{Dynamic Regret Bound of \SafeOGD}
We prove the following regret bound for \SafeOGD.
\begin{theorem}[Dynamic Regret Bound of \SafeOGD]\label{theorem:SafeOGD_Control}
Consider the \SafeNSC problem. \SafeOGD achieves against any sequence of comparators $(K^\star_1, \dots, K^\star_T) \in \calK_1 \times \cdots \times \calK_T$,
\begin{equation}
    \DyRegNSC \leq \frac{\eta T G_f^2}{2} + \frac{7 D_f^2}{4 \eta} + \frac{D_f C_T}{\eta} + \frac{D_f S_T}{\eta},
    \label{eq:theorem_SafeOGD_bound_init}
\end{equation}
where $G_f \triangleq G D d_x d_u (\kappa_{B}+1)$, $D_f\triangleq2\kappa\sqrt{d}$, $D\triangleq \max \{ \frac{W}{\gamma}, \frac{W\kappa}{\gamma}\}$, and $d\triangleq\min \left\{d_{u}, d_{x}\right\}$.

Specifically, for $\eta=\calO\left(\frac{1}{\sqrt{T}}\right)$, we have
\begin{equation}
    \DyRegNSC \leq \calO\left(\sqrt{T}\left(1+C_{T}+S_{T}\right)\right).
   	\label{eq:theorem_SafeOGD_bound}
\end{equation}
\end{theorem}

\noindent\textit{Proof:}
By convexity of $f_t$, we have
\begin{equation}
    \begin{aligned}
        & f_t\left(K_t\right)-f_t\left(K^\star_t\right) \\
         \leq & \left\langle\nabla f_t\left(K_t\right), K_t-K^\star_t\right\rangle \\
        = & \frac{1}{\eta}\left\langle K_t-K_{t+1}^{\prime}, K_t-K^\star_t\right\rangle \\
        = & \frac{1}{2 \eta}\left(\left\|K_t-K^\star_t\right\|_{\mathrm{F}}^2-\left\|K_{t+1}^{\prime}-K^\star_t\right\|_{\mathrm{F}}^2+\left\|K_t-K_{t+1}^{\prime}\right\|_{\mathrm{F}}^2\right) \\
        = & \frac{1}{2 \eta}\left(\left\|K_t-K^\star_t\right\|_{\mathrm{F}}^2-\left\|K_{t+1}^{\prime}-K^\star_t\right\|_{\mathrm{F}}^2\right)+\frac{\eta}{2}\left\|\nabla f_t\left(K_t\right)\right\|_{\mathrm{F}}^2 \\
        \leq & \frac{1}{2 \eta}\left(\left\|K_t-K^\star_t\right\|_{\mathrm{F}}^2-\left\|\bar{K}_{t+1}-K^\star_t\right\|_{\mathrm{F}}^2\right)+\frac{\eta}{2} G_f^2 ,
        \label{eq:theorem_SafeOGD_1}
    \end{aligned}
\end{equation}
where the last inequality holds due to the Pythagorean theorem \cite{hazan2016introduction} and \Cref{lemma:Gradient}.
Consider now the term $\left\|\bar{K}_{t+1}-K^\star_t\right\|_{\mathrm{F}}^2$: 
\begin{equation}
    \begin{aligned}
        \left\|\bar{{K}}_{t+1}-{K}_t^\star\right\|_{\mathrm{F}}^2 &= \left\|{{K}}_{t+1}-{K}_t^\star\right\|_{\mathrm{F}}^2 +\left\|{K}_{t+1}-\bar{{K}}_{t+1}\right\|_{\mathrm{F}}^2 \\ 
        &\quad - 2 \left\langle{{K}}_{t+1}-{K}_t^\star, {{K}}_{t+1}-\bar{{K}}_{t+1}\right\rangle.
        \label{eq:theorem_SafeOGD_2}
    \end{aligned}
\end{equation}
Substituting \cref{eq:theorem_SafeOGD_2} into \cref{eq:theorem_SafeOGD_1} gives
\begin{equation}
    \begin{aligned}
        &{f}_{t}\left(K_{t}\right)-{f}_{t}\left(K_{t}^{*}\right) \\
        \leq & \frac{1}{2 \eta}\left(\left\|{K}_t-{K}_t^\star\right\|_{\mathrm{F}}^2 - \left\|{{K}}_{t+1}-{K}_t^\star\right\|_{\mathrm{F}}^2 - \right. \\
        &\left. \left\|{K}_{t+1}-\bar{{K}}_{t+1}\right\|_{\mathrm{F}}^2 + 2 \left\langle{{K}}_{t+1}-{K}_t^\star, {{K}}_{t+1}-\bar{{K}}_{t+1}\right\rangle\right)+\frac{\eta}{2} G_f^2 \\
        \leq & \frac{1}{2 \eta}\left(\left\|{K}_t-{K}_t^\star\right\|_{\mathrm{F}}^2 - \left\|{{K}}_{t+1}-{K}_t^\star\right\|_{\mathrm{F}}^2 - \left\|{K}_{t+1}-\bar{{K}}_{t+1}\right\|_{\mathrm{F}}^2 \right.\\
        & \ \left.+ 2 \left\|{{K}}_{t+1}-{K}_t^\star\right\|_{\mathrm{F}} \left\|{{K}}_{t+1}-\bar{{K}}_{t+1}\right\|_{\mathrm{F}} \right)+\frac{\eta}{2} G_f^2 \\
        \leq& \frac{1}{2 \eta}\left(\left\|{K}_t-{K}_t^\star\right\|_{\mathrm{F}}^2 - \left\|{{K}}_{t+1}-{K}_t^\star\right\|_{\mathrm{F}}^2 \right) + \frac{D_f \zeta_t}{\eta} +\frac{\eta}{2} G_f^2 \\
        =& \frac{1}{2 \eta}\left(\left\|{K}_t\right\|_{\mathrm{F}}^2 - \left\|{{K}}_{t+1}\right\|_{\mathrm{F}}^2 \right) +\frac{1}{\eta} \left\langle{K}_{t+1}-{K}_t, {K}_t^\star\right\rangle \\
        &\ + \frac{D_f \zeta_t}{\eta} +\frac{\eta}{2} G_f^2,
        \label{eq:theorem_SafeOGD_3}
    \end{aligned}
\end{equation}
where the second inequality holds due to the Cauchy-Schwarz inequality, and the third inequality holds due to $\left\|K_{t+1}-\bar{K}_{t+1}\right\|_{\mathrm{F}}^2 \geq 0$, $\left\|{K}_{t+1}-K^\star_t\right\|_{\mathrm{F}} \leq D_f$ by \Cref{lemma:K_norm}, and $\zeta_t \triangleq \left\|\bar{K}_{t+1} - {K}_{t+1} \right\|_{\mathrm{F}}$ by definition.

Summing \cref{eq:theorem_SafeOGD_3} over all iterations, we have for any comparators sequence $(K^\star_1, \dots, K^\star_T) \in \calK_1 \times \cdots \times \calK_T$ that
\begin{equation}
    \begin{aligned}
        &\sum_{t=1}^T f_t\left(K_t\right) - \sum_{t=1}^T f_t\left(K^\star_t\right) \\
        \leq& \frac{1}{2 \eta}\left\|K_1\right\|_{\mathrm{F}}^2 - \frac{1}{2 \eta}\left\|K_{T+1}\right\|_{\mathrm{F}}^2 + \frac{1}{\eta} \sum_{t=1}^T\langle K_{t+1}-K_t, K^\star_t \rangle\\
        & \ + \frac{D_f}{\eta} \sum_{t=1}^T\zeta_t +\frac{\eta T}{2} G_f^2 \\
        =& \frac{1}{2 \eta}\left\|K_1\right\|_{\mathrm{F}}^2- \frac{1}{2 \eta}\left\|K_{T+1}\right\|_{\mathrm{F}}^2 + \frac{1}{\eta}\left(\langle K_{T+1}, K^\star_T \rangle - \langle K_1, K^\star_1 \rangle \right) \\
        & \ +\frac{1}{\eta} \sum_{t=2}^T \langle K^\star_{t-1}-K^\star_t, K_t \rangle + \frac{D_f}{\eta} \sum_{t=1}^T\zeta_t + \frac{\eta T}{2} G_f^2 \\
        \leq& \frac{1}{2 \eta}\left\|K_1\right\|_{\mathrm{F}}^2 + \frac{1}{\eta}\left(\langle K_{T+1}, K^\star_T \rangle - \langle K_1, K^\star_1 \rangle \right) \\
        & \ +\frac{1}{\eta} \sum_{t=2}^T \langle K^\star_{t-1}-K^\star_t, K_t \rangle  + \frac{D_f}{\eta} \sum_{t=1}^T\zeta_t + \frac{\eta T}{2} G_f^2 \\
        \leq& \frac{7 D_f^2}{4 \eta} + \frac{D_f}{\eta} C_T + \frac{D_f}{\eta} S_T + \frac{\eta T}{2} G_f^2,
        \label{eq:theorem_SafeOGD_4}
    \end{aligned}
\end{equation}
where the last step holds due to \Cref{lemma:K_norm} and the Cauchy-Schwarz inequality, \ie $\left\|K_1\right\|_{\mathrm{F}}^2 \leq D_f^2$, $\langle K_{T+1}, K^\star_T \rangle \leq \left\|K_{T+1}\right\|_{\mathrm{F}}\left\|K^\star_T\right\|_{\mathrm{F}} \leq D_f^2$, $-\langle K_1, K^\star_1 \rangle \leq \frac{1}{4}\left\|K_1-K^\star_1\right\|_{\mathrm{F}}^2 \leq \frac{1}{4} D_f^2$, $\langle K^\star_{t-1}-K^\star_t, K_t \rangle  \leq\left\|K^\star_{t-1}-K^\star_t\right\|_{\mathrm{F}}\left\|K_t\right\|_{\mathrm{F}} \leq D_f\left\|K^\star_{t-1}-K^\star_t\right\|_{\mathrm{F}}$, along with the definitions of path length $C_T$ and set variation $S_T$.
\qed

The dependency on $C_T$ results from the time-varying sequence of comparators.
Specifically, any optimal dynamic regret bound for \OCO is $\Omega\left(\sqrt{T(1+C_T)}\right)$, and thus the bound necessarily depends on $C_T$ in the worst case \cite{zhang2018adaptive}. 

The dependency on $S_T$ results from the domain sets being time-varying. $S_T$ is zero when the domain sets are time-\underline{in}variant (\Cref{remark:bound_reduction}).  Thus, $S_T$ can be sublinear in decision-making applications where any two consecutive safe sets differ a little (\eg in high-frequency control applications where the control input is updated every a few tenths of milliseconds, then, the safety set may change only a little between consecutive time steps).

\subsection{Regret Bounds in the Time-\underline{In}variant Domain Case}\label{subsec:invariant}
When the domain set is time-\underline{in}variant, the regret bounds in \cref{eq:theorem_SafeOGD_bound} reduce to the results in the standard \OCO setting

\begin{remark}[Regret Bounds in the Time-\underline{In}variant Domain Case]\label{remark:bound_reduction}
    When the domain set is time-\underline{in}variant, \ie $\calK_1 = \dots = \calK_T$, we have $S_T = 0$ by definition. Hence, the dynamic regret bound in \cref{eq:theorem_SafeOGD_bound} reduces to $\calO\left(\sqrt{T}\left(1+C_{T}\right)\right)$, \ie it becomes equal to the dynamic regret bound of \OGD in the standard \OCO setting  \cite{zinkevich2003online}.
\end{remark}
\section{Numerical Evaluations}\label{sec:exp}
We compare \SafeOGD with the safe $H_2$ and $H_\infty$ controllers in simulated scenarios of safe control of a quadrotor aiming to stay at a hovering position. {We implement $H_2$ and $H_\infty$ controllers based on \cite[eqs. (2.15) \& (2.19)]{anderson2019system} and use \cite[
Theorem~3]{martin2022safe} to account for safety constraints.} We implement $H_2$ and $H_\infty$ with three different horizons, \ie $N = 1, \ 5, \ 10$. Supplementary numerical experiments, that compare \SafeOGD with \OCOM controllers \cite{agarwal2019online}, are presented in \Cref{app:supp_exp}. 
Our code is open-sourced at: \href{https://github.com/UM-iRaL/Non-Stochastic-Control}{https://github.com/UM-iRaL/Non-Stochastic-Control}.


\myParagraph{Tested Noise Types}
We corrupt the system dynamics with diverse noise drawn for the Gaussian, Uniform, Gamma, Beta, Exponential, or Weibull distribution. 

\myParagraph{Simulation Setup} We consider a quadrotor model with state vector its position and velocity, and control input its roll, pitch, and total thrust. The quadrotor's goal is to stay at a predefined hovering position.  To this end, we focus on its linearized dynamics, taking the form 
\begin{equation}
    \begin{aligned}
        x_{t+1} = A x_{t} + B u_{t} + w_{t}
    \end{aligned}
\end{equation}
where
\addtolength{\arraycolsep}{-1.75pt}
\begin{equation*}
    \begin{aligned} 
            A\!&=\!\left[\begin{array}{cccccc}
            1 & 0 & 0 & 0.1 & 0 & 0 \\
            0 & 1 & 0 & 0 & 0.1 & 0 \\
            0 & 0 & 1 & 0 & 0 & 0.1 \\
            0 & 0 & 0 & 1 & 0 & 0 \\
            0 & 0 & 0 & 0 & 1 & 0 \\
            0 & 0 & 0 & 0 & 0 & 1 
            \end{array}\right], B\!=\!\left[\begin{array}{ccc}
            -\frac{4.91}{100} & 0 & 0 \\
             0 & \frac{4.91}{100} & 0 \\
             0 & 0 & \frac{1}{200} \\
             -\frac{98.1}{100} & 0	& 0 \\
             0 & \frac{98.1}{100} & 0 \\
             0 & 0 & \frac{1}{10}
            \end{array}\right].
    \end{aligned}
\end{equation*}
\addtolength{\arraycolsep}{1.75pt}

We choose the safety constraints: 
\begin{equation}
    \begin{aligned}
        -\textbf{1}_{6\times1} & \leq  x_{t} \leq  \textbf{1}_{6\times1}, \\
        [-\pi \ -\pi \ -20]^\top & \leq u_{t} \leq  [\pi \ \pi \ 20]^\top,
        \label{eq:exp_safety}
    \end{aligned}
\end{equation}
and we assume noise such that $\| w_{t} \|\, \leq 0.1$ for all $t$. 

We consider that the loss functions take the form of $c_t(x_{t+1}, u_t) = x_{t+1}^\top x_{t+1} + u_{t}^\top u_{t}$. 

We simulate the setting for $T~=~500$ time steps.

\begin{remark}[Time-Varying Domain Set]
The domain set $\calK_t$ for the quadrotor system is \textit{time-varying} even though the safety constraints in \cref{eq:exp_safety} are time-invariant, since $\calK_t$ depends on the time-varying state $x_t$ over $T$ in \cref{eq:lemma_calK_t_2}.
\end{remark}

{
\myParagraph{Summary of Results} The simulation results are presented in \Cref{table_UAV_loss_H5} (cumulative loss performance) and \Cref{table_UAV_time_H5} (running time). All methods ensure the safety constraints in \cref{eq:exp_safety} are satisfied. \Cref{alg:SafeOGD_Control} 
demonstrates better performance in comparison to ${H}_2$ and ${H}_\infty$ with $N=1$ and $N=5$ in terms of cumulative loss across the tested types of noise. 
${H}_2$ and ${H}_\infty$ with $N=10$ incur lower cumulative loss than \Cref{alg:SafeOGD_Control}. However, as shown in \Cref{table_UAV_time_H5}, \Cref{alg:SafeOGD_Control} is computationally more efficient. Specifically, \Cref{alg:SafeOGD_Control} is $9$ and $114$ times faster than ${H}_2$ and ${H}_\infty$ with $N=10$ on average, respectively.
}

\renewcommand{\arraystretch}{1.2} 
\begin{table}[t]
  \centering
     \caption{Comparison of \SafeOGD with the safe $H_2$ and $H_\infty$ controllers in terms of cumulative loss over.
     }
     \label{table_UAV_loss_H5}
     \resizebox{\columnwidth}{!}{
     {
     \begin{tabular}{cccccccc}
     \toprule
     \multirow{2}{*}{Noise Distribution} & \multirow{2}{*}{Ours} & \multicolumn{2}{c}{$N=1$} & \multicolumn{2}{c}{$N=5$}	& \multicolumn{2}{c}{$N=10$} \cr
     \cmidrule(lr){3-4} \cmidrule(lr){5-6} \cmidrule(lr){7-8} & &  $H_2$ & $H_\infty$ & $H_2$ & $H_\infty$ & $H_2$ & $H_\infty$\cr
    \midrule
	 Gaussian   &  44.05   & 61.81 & 93.44 &  47.96 &  52.03 & 30.66 & 48.69 \cr
	 Uniform   &  151.49 & 724.98 & 1859.61 &  331.32  &  323.42 & 100.21 & 53.86 \cr
	 Gamma   &  159.21 & 811.09 & 2082.12 &   372.52  &  364.26 & 112.90 & 60.77 \cr
	 Beta   &  186.98 & 836.41 & 2152.63 & 386.30  &  375.73 & 116.70 & 62.40 \cr
	 Exponential  & 126.69 & 552.73 & 1421.90 &   259.82  &  250.76 & 79.25 & 44.35 \cr
	 Weibull   &  195.71 & 873.09 & 2246.31 &   405.70  &  392.94 & 122.63 & 65.86 \cr
    \midrule
	 \textbf{Average}   &  142.50 & 643.35 & 1642.67 &  300.60  &  293.19 & 93.72 & 55.99 \cr
	 \textbf{Standard Deviation}  &  53.92 & 307.00 & 814.06 &  134.16  &  128.60 & 34.53 & 8.43 \cr
    \bottomrule
    \end{tabular}}
     }
\end{table}


\renewcommand{\arraystretch}{1.2} 
\begin{table}[t]
  \centering
     \caption{Comparison of  \SafeOGD \Cref{alg:SafeOGD_Control} with the safe $H_2$ and $H_\infty$ controllers in terms of computation time in seconds. 
     }
     \label{table_UAV_time_H5}
     \resizebox{\columnwidth}{!}{
     {
     \begin{tabular}{cccccccc}
     \toprule
     \multirow{2}{*}{Noise Distribution} & \multirow{2}{*}{Ours} & \multicolumn{2}{c}{$N=1$} & \multicolumn{2}{c}{$N=5$}	& \multicolumn{2}{c}{$N=10$} \cr
     \cmidrule(lr){3-4} \cmidrule(lr){5-6} \cmidrule(lr){7-8} & &  $H_2$ & $H_\infty$ & $H_2$ & $H_\infty$ & $H_2$ & $H_\infty$\cr
    \midrule
	 \textbf{Average}   &  0.1484 & 0.3712 & 0.6429 &  0.6033  &  1.3693 & 1.3854 & 17.0248 \cr
	 \textbf{Standard Deviation}  &  0.0342 & 0.0143 & 0.0116 & 0.0282  &  0.2741 & 0.0673 & 0.3691 \cr
    \bottomrule
    \end{tabular}}
     }
\end{table}

\section{Conclusion} \label{sec:con}

We studied the problem of \textit{Safe Non-Stochastic Control of Linear Dynamical Systems} (\Cref{prob:SafeNSC}), and provided the \SafeOGD algorithm that guarantees (i) zero constraint violation of convex time-varying constraints,  and (ii) bounded {dynamic regret} against any linear time-varying control policy with safety guarantees (\Cref{theorem:SafeOGD_Control}). 
We demonstrated that the dynamic regret bound of \SafeOGD reduces to that in the standard OCO setting \cite{zinkevich2003online} when the optimization domain is time-\underline{in}variant (\Cref{remark:bound_reduction}).

We evaluated our algorithm in simulated scenarios of safe control of a quadrotor aiming to maintain a hovering position in the presence of unpredictable disturbances.  {We observed that the \SafeOGD-based controller achieved comparable cumulative loss and better computational time compared to safe $H_2$ and $H_\infty$ controllers \cite{anderson2019system,martin2022safe}. }

\myParagraph{Future Work} We will investigate the optimality of the regret bound of the \SafeOGD algorithm. We will also investigate conditions for the recursive feasibility of time-varying domain set $\calK_t$. Further, we will apply the algorithm to real-world robotic systems (quadrotors) to demonstrate resilient online control against unpredictable wind. To this end, we will extend the algorithms to nonlinear systems.



\bibliographystyle{IEEEtran}
\bibliography{References}

\appendix


\myParagraph{Notation}
We denote $\|\cdot\|$ as $2$-norm for vectors and $\mathrm{op}$-norm for matrices. We use $\|\cdot\|_{\mathrm{F}}$ as Frobenius norm.

\renewcommand{\arraystretch}{1.2} 
\begin{table*}[!]
  \centering
     \caption{Comparison of the \SafeOGD and \DAC \cite{agarwal2019online} controllers with two step sizes in terms of cumulative loss for $1000$ time steps ---the \blue{blue} numbers correspond to the \blue{best} performance and the \red{red} numbers correspond to the \red{worse}.}
     \label{table_OCOvsOCOM_single}
     \resizebox{1.8\columnwidth}{!}{
     {
     \begin{tabular}{ccccccccc}
     \toprule
 	 \multirow{3}{*}{Noise Distribution} & \multicolumn{4}{c}{Sinusoidal Weights (\cref{eq_loss_1})} & \multicolumn{4}{c}{Step Weights (\cref{eq_loss_2})}		\cr
    \cmidrule(lr){2-5} \cmidrule(lr){6-9} & \multicolumn{2}{c}{Ours}  & \multicolumn{2}{c}{{\fontsize{7.2}{7.2}\selectfont\sf{\textbf{DAC}}}}  & \multicolumn{2}{c}{Ours}  & \multicolumn{2}{c}{{\fontsize{7.2}{7.2}\selectfont\sf{\textbf{DAC}}}}  \cr
     &  $\eta_1$ & $\eta_2$  & $\eta_1$ & $\eta_2$ & $\eta_1$ & $\eta_2$ &  $\eta_1$ & $\eta_2$  \cr
    \midrule
	 Gaussian & 1769 & 1732  & \red{1838} & \blue{1561}   & 952 & 913  & \red{991} & \blue{860}    \cr
	 Uniform & \red{2839} & 2822  & 2649 & \blue{2428}  & \red{1555} & 1538  & 1508 & \blue{1352}  \cr
	 Gamma & 845 & \blue{690}  & \red{30323} & 8193  & 591 & \blue{423}  & \red{29252} & 4746 	 \cr
	 Beta & \red{3518} & 3494  & 3045 & \blue{2628}  & \red{1921} & 1899 &  1795 & \blue{1489}   \cr
	 Exponential & 1359 & \blue{1252} & \red{54470} & 20273  & 866 & \blue{726}  & \red{44821} & 9122  \cr
	 Weibull & 1732 & \blue{1540}  & \red{73100} & 7271  & 1332 & \blue{1118}  & \red{72005} & 4776  \cr
    \midrule
	 \textbf{Average} &  2010 & \blue{1922} & \red{27571} & 7059  & 1203 & \blue{1103} & \red{25062} & 3724 \cr
	 \textbf{Standard Deviation} &  \blue{988} & 1042  & \red{30623} & 7032  & \blue{491} & 541  & \red{29282} & 3166 \cr
    \bottomrule
    \end{tabular}}
     }
\end{table*}


\subsection{Supporting Lemmas}\label{app:lemma_SafeOGD_control}

\begin{lemma}[Bounded State and Control]\label{lemma:bounded_variables}
Let $K_t$ with $\|K_t\|\, \leq \kappa$ be the stable linear controllers at each iteration $t \in \TimeSeq$, \ie $\| A_t - B_t K_t \| \,\leq 1-\gamma$. Suppose the initial state is $x_1=0$. Define $D\triangleq \max \{ \frac{W}{\gamma}, \frac{W\kappa}{\gamma}\}$. Then, we have
\begin{equation}
	\left\|x_{t}\right\|  \leq D, \ \left\|u_{t}\right\| \leq D, \ \forall t \in \TimeSeq 
\end{equation}
\end{lemma}
\noindent\textit{Proof:}
By definition, the state propagated by the sequence of time-varying controller $K_{1}, \dots, K_{t}$ is 
\begin{equation}
	x_{t+1} = \sum_{i=0}^{t-1} \widetilde{A}_{K_{t:t-i+1}} w_{t-i}.
\end{equation}
where $\widetilde{A}_{K^{*}_{t:t-i}} \triangleq \prod_{\tau=t}^{t-i}\left(A_\tau-B_\tau K^{*}_\tau\right)$ and  $\widetilde{A}_{K^{*}_{t:t-i}} \triangleq \myIdentity$ if $i~<~0$.
Hence, we have
\begin{equation}
    \begin{aligned}
        \| x_{t+1} \| &= \| \sum_{i=0}^{t-1} \widetilde{A}_{K_{t:t-i+1}} w_{t-i} \|\, \leq  \sum_{i=0}^{t-1} \| \widetilde{A}_{K_{t:t-i+1}} w_{t-i} \| \\
        &\leq  W \sum_{i=0}^{t-1} \| \,\widetilde{A}_{K_{t:t-i+1}} \|\, \leq  W \sum_{i=0}^{t-1} (1-\gamma)^i \\       
        &= W \frac{1-(1-\gamma)^t}{\gamma},
    \end{aligned}
\end{equation}
which implies $\| x \| \, \leq \frac{W}{\gamma}$ for all $t \in \TimeSeq$.

Consider the control input, we have
\begin{equation}
    \| u_t \| = \| -K_t x_t \| \, \leq \kappa \frac{W}{\gamma}.
\end{equation}
\qed
\begin{lemma}[Bounded Gradient]\label{lemma:Gradient} 
Define $D\triangleq \max \{ \frac{W}{\gamma}, \frac{W\kappa}{\gamma}\}$. The loss $f_{t}: \mathbb{R}^{d_u \times d_x} \rightarrow \mathbb{R}$ has bounded gradient norm $G_f$, \ie $\left\|\nabla_{K} {f}_{t}(K)\right\|_{\mathrm{F}} \leq G_{f}$ holds for any $K \in \mathcal{K}_t$ and any $t \in \TimeSeq$, where $G_{f} \leq G D d_x d_u (\kappa_{B}+1)$.
\end{lemma}

\noindent\textit{Proof:}
We need to bound $\nabla_{K_{p, q}} {f}_{t}(M)$ for every $p \in\{1, \dots,d_{u}\}$ and $q \in\{1, \dots,d_{x}\}$,
\begin{equation}
	\left|\nabla_{K_{p, q}} {f}_{t}(K)\right| \leq G\left\|\frac{\partial x_{t+1}(K)}{\partial K_{p, q}}\right\|_{\mathrm{F}} + G\left\|\frac{\partial u_{t}(K)}{\partial K_{p, q}}\right\|_{\mathrm{F}} .
\end{equation}

Now we aim to bound the two terms on the right-hand side respectively:
\begin{equation}
	\begin{aligned}
        \left\|\frac{\partial x_{t+1}(K)}{\partial K_{p, q}}\right\|_{\mathrm{F}}  &= \left\|\frac{\partial \left( A_t x_{t} - B_t K x_t + w_t \right)}{\partial K_{p, q}}\right\|_{\mathrm{F}} = \left\|\frac{\partial  B_t K x_t }{\partial K_{p, q}}\right\|_{\mathrm{F}} \\
        &\leq \kappa_B D \left\|\frac{\partial K }{\partial K_{p, q}}\right\|_{\mathrm{F}} = \kappa_B D, \\
        \left\|\frac{\partial u_{t}(K)}{\partial K_{p, q}}\right\|_{\mathrm{F}} &= \left\|\frac{\partial \left(- K x_t\right) }{\partial K_{p, q}}\right\|_{\mathrm{F}} = \left\|\frac{\partial  K x_t }{\partial K_{p, q}}\right\|_{\mathrm{F}} \\
        &\leq  D \left\|\frac{\partial K }{\partial K_{p, q}}\right\|_{\mathrm{F}} =  D.
	\end{aligned}
\end{equation}

Therefore, we have
\begin{equation}
    \left|\nabla_{K_{p, q}} {f}_{t}(K)\right| \leq G \kappa_{B} D + G D = G D (\kappa_{B}+1).
\end{equation}
Thus, $\left\|\nabla_{K} {f}_{t}(K)\right\|_{\mathrm{F}}$ is at most $G D d_x d_u (\kappa_{B}+1)$.
\qed

\begin{lemma}[Bounded Domain of Control Gain]\label{lemma:K_norm} 
For any $K_1, K_2 \in \calK \subset \mathbb{R}^{d_{u} \times d_{x}}$, where $\calK\triangleq\{K \mid \|K\|\,\leq \kappa\}$, we have $\| K_1 - K_2\|_{\mathrm{F}} \, \leq D_f$, where $D_f\triangleq2\kappa\sqrt{d}$ and $d\triangleq\min \left\{d_{u}, d_{x}\right\}$.
\end{lemma}

\noindent\textit{Proof:} For any matrix $X \in \mathbb{R}^{m \times n}$,
\begin{equation}
	\|X\| \,\leq\|X\|_{\mathrm{F}} \, \leq \sqrt{\min\{m,n\}}\|X\|.	
\end{equation}

Therefore, 
\begin{equation}
	\|K_1 - K_2\|_{\mathrm{F}}\, \leq \sqrt{d}\|K_1 - K_2\|\, \leq \sqrt{d} \left(\|K_1\| + \|K_2\|\right) = 2\kappa\sqrt{d}.	
\end{equation}
\qed

\subsection{Supplementary Numerical Experiments}\label{app:supp_exp}
In this experiment, we compare our algorithm with state-of-the-art \OCOM controller \cite{agarwal2019online}.  We showcase that online optimization with memory does not necessarily result in superior performance.


\myParagraph{Compared Algorithms} 
We compare the \SafeOGD-based controller with the memory-based \DAC \cite{agarwal2019online} controller.

\myParagraph{Simulation Setup} 
We follow a setup similar to \cite{zhao2022non}. We consider linear systems of the form
\begin{equation}\label{eq:aux_exp}
    \begin{aligned}
        x_{t+1} &= A x_{t} + B u_{t} + w_{t},
    \end{aligned}
\end{equation}
where (i) $x_t \in \mathbb{R}^2$, (ii) $u_t\in \mathbb{R}$, and (iii) ${w}_{t}$ and the elements of  $A$ and $B$ are sampled from various distributions, \ie Gaussian, Uniform, Gamma, Beta, Exponential, or Weibull distributions. 
We consider linear time-invariant systems and impose constraints only on the control input.  This induces a time-\underline{in}variant domain set of optimization, as required by the \DAC controller~\cite{agarwal2019online}. 
Specifically, we use the 
control constraint $L_{u} u \leq l_{u}$, \ie $- L_{u} K x_t \leq l_{u}$. If we upper bound  $x_t$ with  upper bound $D^\prime$ achieved by the \OCOM controller \cite[Lemma~5.5]{agarwal2019online}, then the optimization domain in \Cref{lemma:calK_t} becomes time-\underline{in}variant, specifically,
\begin{equation}
    \begin{aligned}
    \calK &\triangleq \{  K \mid -L_{u} D^\prime K  \leq l_{u}, \ \| K \|\; \leq \kappa, \   \| A - B K\| \;\leq 1-\gamma \}.
    \end{aligned}
    \label{eq:lemma_calK_t}
\end{equation}

We compare the \SafeOGD and \DAC controllers across two different step sizes $\eta_1$ and $\eta_2$ {to investigate how the step sizes affect their performance}. {The \DAC controller has a memory length of $10$.}



The loss function has the form $c_t(x_{t},u_t) =  q_t x_t^\top x_t + r_t u_t^\top u_t$, where $q_t$, $r_t \in \mathbb{R}$ are time-varying weights. Particularly, we consider the following two cases:

\begin{enumerate}
    \item Sinusoidal weights defined as 
    \begin{equation}
        q_t = \sin (t/10\pi),\ r_t = \sin (t/20\pi).
        \label{eq_loss_1}
    \end{equation}
    \item Step weights defined as 
    \begin{equation}
        \begin{aligned}
            (q_t, r_t) = \left\{ \begin{array}{cc} \left(\frac{\log(2)}{2}, 1 \right)  , \ & t \leq T/5,  \\ 
            \left(1, 1 \right) , \ & T/5 < t \leq 2T/5, \\
            \left(\frac{\log(2)}{2}, \frac{\log(2)}{2} \right)  , \ & 2T/5 < t \leq 3T/5, \\  
            \left(1, \frac{\log(2)}{2} \right)  , \ & 3T/5 < t \leq 4T/5,  \\
            \left(\frac{\log(2)}{2}, 1 \right)   , \ & 4T/5 < t \leq T.
            \end{array} \right.
        \end{aligned}
        \label{eq_loss_2}
    \end{equation}
\end{enumerate}

\myParagraph{Results} 
The results are summarized in \Cref{table_OCOvsOCOM_single}, showing that \SafeOGD outperforms \DAC in terms of the average and standard deviation of cumulative loss. In more detail, \SafeOGD has comparable performance to \DAC under Gaussian, Uniform, and Beta distributions, and is better under Gamma, Exponential, and Weibull distributions. {We hypothesize that the reason for the latter is that the \DAC controller minimizes a truncated unary loss, instead of the actual loss.} In addition, the performance of \DAC heavily relies on step size tuning, \eg under Gamma and Weibull distributions, {as demonstrated by the large difference in cumulative loss across $\eta_1$ and $\eta_2$}. By contrast, the cumulative loss of \SafeOGD varies less as we change the step size.

\subsection{Discussion on Recursive Feasibility}\label{app:rf}
To ensure recursive feasibility of $\calK_t$, we may utilize a standard approach in robust model predictive control~\cite{mayne2000constrained,mayne2005robust}. The method assumes there exists a sequence of control inputs over a given lookahead horizon $N$ such that the system can be driven into a tightened safe set. Then, this safe set is assumed to be forward invariant by applying a known baseline controller. Finally, the recursive feasibility is guaranteed {by the combination of (i) the last $N-1$ control inputs from the sequence of control at the last iteration, and (ii) the baseline controller; particularly, (i) and (ii) form a feasible sequence of control inputs}. For simplicity in the presentation, we consider the linear time-invariant system\footnote{The discussion generalizes to linear time-varying systems following similar steps by adding time index to matrices $A$, $B$, and $K^s$.}
\begin{equation}
	x_{t+1} = A_{} x_{t} + B_{} u_{t} + w_{t}, \quad t=1, \ldots, T,
	\label{eq:LTI}
\end{equation}
and its nominal noiseless system
\begin{equation}
	\bar{x}_{t+1} = A_{} \bar{x}_{t} + B_{} \bar{u}_{t} , \quad t=1, \ldots, T.
	\label{eq:LTI_n}
\end{equation}

We use the following notations:
\begin{itemize}
    \item $\oplus$ and $\ominus$ is the Minkowski sum and subtraction;
    \item $N$ is the lookahead horizon;
    \item $K^s$ is a known baseline safe controller;
    \item $\calZ$ is a known disturbance invariant set for the system in \cref{eq:LTV}, \ie $(A - B K^s) \calZ \oplus \calW \subseteq \calZ$; 
    \item $\calS_{t+i}$ is the state constraint on $x_{t+i}$, where $i \in \{1,\dots,N\}$; 
    \item $\calU_{t+j}$ is the control input constraint on $u_{t+j}$, where $j \in \{0,\dots,N-1\}$;
    \item $\bar{\calS}_{t+i} \triangleq \calS_{t+i} \ominus \calZ$ such that $\bar{x}_t \in \bar{\calS}_{t+i}$ implies ${x}_t \in {\calS}_{t+i}$; 
    \item $\bar{\calU}_{t+j} \triangleq \calU_{t+j} \ominus K^s \calZ$ such that $\bar{u}_t \in \bar{\calU}_{t+j}$ implies ${u}_t \in {\calU}_{t+j}$;
    \item $\calS_{f}$ is a terminal set, defined in \Cref{ass:terminal_cond} to enable recursive feasibility.

\end{itemize}

We assume the safety constraints over the lookahead horizon $N$ are known.
\begin{assumption}[Future Information]
    We assume that the safety constraints over the lookahead horizon $N$, \ie $\calS_{t+i}$ and $\calU_{t+j}$, where $i \in \{1,\dots,N\}$ and $j \in \{0,\dots,N-1\}$, are known at iteration $t$.
\end{assumption}

To achieve recursive feasibility, we have the following assumption on the terminal set ${\calS}_{f}$ and the baseline safe controller $K^s$.

\begin{assumption}[Terminal Condition]\label{ass:terminal_cond}
    We assume that, at each iteration $t$, the terminal set ${\calS}_{f}$ and the baseline safe controller $K^s$ satisfy
    \begin{enumerate}
        \item $\calS_{f} \subset \bar{\calS}_{t+N}$;
        \item $(A - B K^s) {\calS}_{f} \subset {\calS}_{f}$;
        \item $K^s {\calS}_{f} \subset \bar{\calU}_{t+N}$;
        \item $\bar{\calS}_{t+N} \subseteq \bar{\calS}_{t+N+1}$;
        \item $\bar{\calU}_{t+N} \subseteq \bar{\calU}_{t+N+1}$;
        \item $\| K^s \|\; \leq \kappa$ and $ \| A - B K^s\| \;\leq 1-\gamma$. 
    \end{enumerate}
\end{assumption}

The first three conditions are standard assumptions and imply that the baseline safe controller $K^s$ renders $\bar{x}_{t+N+1} = A_{} \bar{x}_{t+N} + B_{} \bar{u}_{t+N} \in \bar{\calS}_{t+N}$ with $\bar{u}_{t+N} = -K^s \bar{x}_{t+N} \in \bar{\calU}_{t+N}$. The fourth and fifth conditions are imposed to handle the time-varying safety constraints and imply that  $\bar{x}_{t+N+1} \in \bar{\calS}_{t+N+1} $ and $\bar{u}_{t+N+1} \in \bar{\calU}_{t+N+1} $, \ie the safety constraints at $t+N+1$ are satisfied by applying the baseline safe controller.

\begin{lemma}[Set of Control Gains that Guarantee Safety and Recursive Feasibility]\label{lemma:calK_t_rf}
Assume that, at iteration $t=1$, there exists a sequence $\{K_1, \dots, K_{N-1}\}$ such that $\bar{x}_{t+i} \in \bar{\calS}_{t+i}, \ \bar{x}_{t+N} \in {\calS}_{f}, \ \bar{u}_{t+j} \in \bar{\calU}_{t+j}, \ \| K_j \|\; \leq \kappa, \  \| A - B K_j\| \;\leq 1-\gamma$, where $i \in \{1,\dots,N-1\}$ and $j \in \{0,\dots,N-1\}$.
Then by choosing $K_t \in \calK_t$, where 
\begin{equation}
    \begin{aligned}
    \calK_t \triangleq \{ & K_t \mid \bar{x}_{t+i} \in \bar{\calS}_{t+i},  \bar{x}_{t+N} \in {\calS}_{f}, \\
     &  \bar{u}_{t+j} \in \bar{\calU}_{t+j}, \  \| K_{t+j} \|\; \leq \kappa, \   \| A - B K_{t+j}\| \;\leq 1-\gamma, \\
     & i \in \{1,\dots,N-1\}, \ j \in \{0,\dots,N-1\}\},
    \end{aligned}
    \label{eq:lemma_calK_t_rf}
\end{equation}
then {$\{K_t, \dots, K_{t+N-1}\}$ is a feasible control sequence}, at each time step $t$ $x_{t+1} \in \calS_{t+1}$ and  $u_t \in \calU_t$, and the recursive feasibility of $\calK_t$ is guaranteed.
\end{lemma}

\noindent\textit{Proof:} The proof follows similarly as in \cite{mayne2000constrained,mayne2005robust}. 
\qed

\begin{remark}[Non-Convexity of $\calK_t$ and Dynamic Regret Guarantee]
    Due to the lookahead horizon $N$, the domain set $\calK_t$ in \Cref{lemma:calK_t_rf} is non-convex in $K_t, \dots, K_{N-1}$. \Cref{alg:SafeOGD_Control} can still be applied for \SafeNSC.  However, \Cref{theorem:SafeOGD_Control} only holds around the neighborhood of the $K_{t}$.  Specifically, \Cref{theorem:SafeOGD_Control} only holds for the sequences of comparators $(K^\star_1, \dots, K^\star_T) \in \tilde{\calK}_1 \times \cdots \times \tilde{\calK}_T$ where each $\tilde{\calK}_t$ is a convex subset of the non-convex set ${\calK}_t$ in \cref{eq:lemma_calK_t_rf}.
\end{remark}

\subsection{Safe Online Convex Optimization with Time-Varying Constraints (\SafeOCO)}\label{app:safeoco}

\begin{figure}[t]
    \centering
    \includegraphics[width=0.49\textwidth]{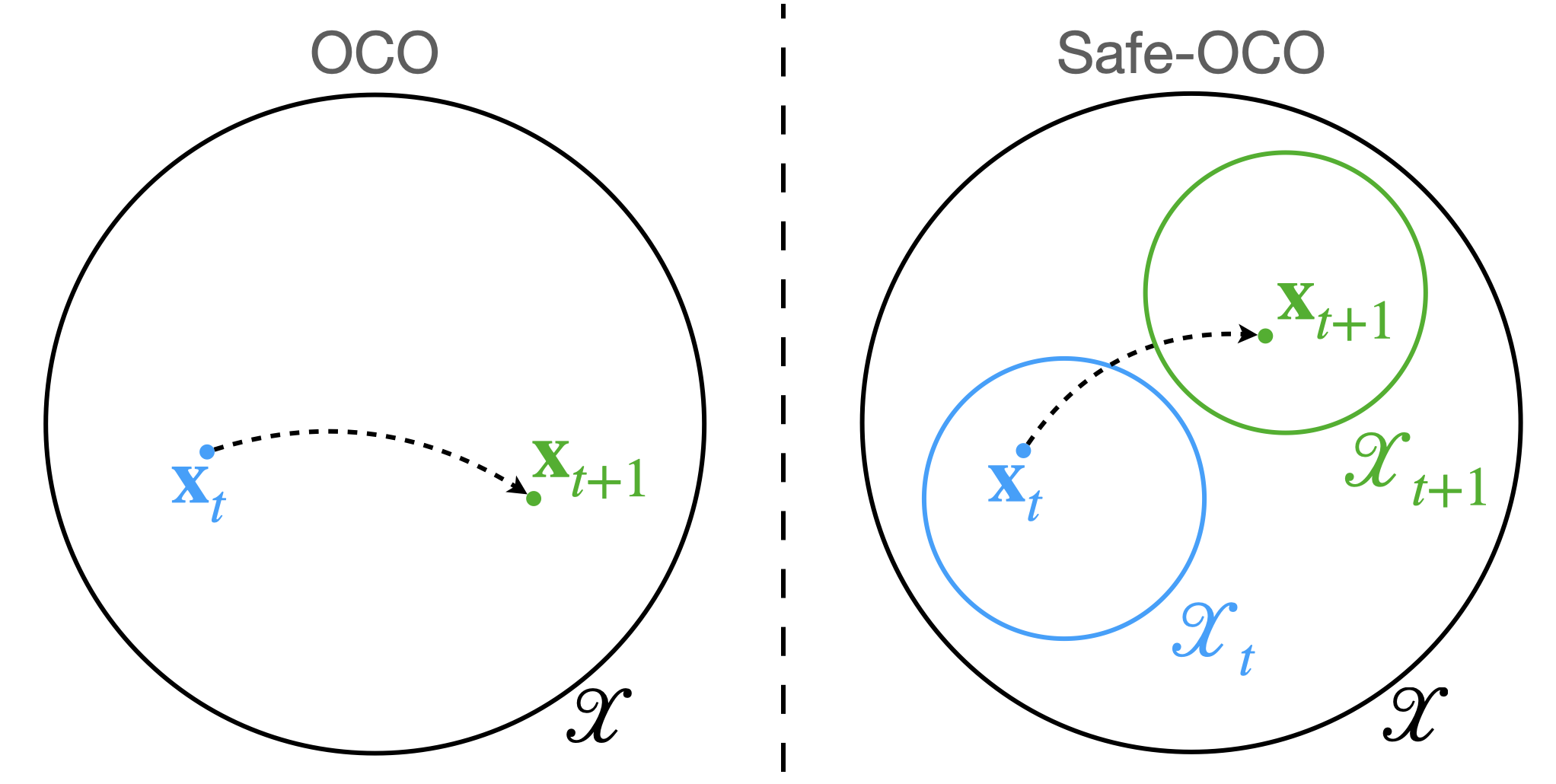}
    \caption{\textbf{Illustration of difference between \OCO and \SafeOCO}. In \OCO, the optimizer  chooses decisions $\myx_t$ and $\myx_{t+1}$ from the same time-\underline{in}variant domain set $\calX$, for all $t=1,\ldots, T$. In \SafeOCO instead, the optimizer chooses decisions from time-varying domain sets, \ie $\myx_{t} \in \calX_{t}$ and $\myx_{t+1} \in \calX_{t+1}$, where $\calX_t$ and $\calX_{t+1}$ are potentially disjoint.}
    \label{fig:OCOvsSafeOCO}
\end{figure} 

We define the problem of \textit{Safe Online Convex Optimization with Time-Varying Constraints} (\Cref{prob:safe_OCO}) for the general online learning problem, along with standard convexity assumptions that we adopt for its solution. This section is of independent interest.

\begin{problem}[Safe Online Convex Optimization with Time-Varying Constraints (\SafeOCO)]\label{prob:safe_OCO}
Two players, an online optimizer and an adversary, choose decisions sequentially over a time horizon $T$. At each time step $t=1,\ldots, T$, the optimizer first chooses a decision $\mathbf{x}_t$ from a known convex set $\calX_t$; then, the adversary chooses a loss $f_t$ to penalize the optimizer's decision. Particularly, the adversary reveals $f_t$ to the optimizer and the optimizer computes its loss $f_t(\myx_t)$.
The optimizer aims to minimize $\SumOneT f_t(\myx_t)$.
\end{problem}

The challenges in solving \SafeOCO, \ie in minimizing $\SumOneT f_t(\myx_t)$, are two: first,  the optimizer gets to know $f_t$ only after $\myx_t$ has been chosen, instead of before; and second, the optimizer must choose $\myx_t$ from a time-varying domain set $\calX_t$, instead of a time-invariant set, where, additionally, $\calX_{t-1}$ is possibly disjoint from $\calX_t$ (\Cref{fig:OCOvsSafeOCO}). Despite the above challenges, we aim to develop an online algorithm for \SafeOCO with sublinear dynamic regret.  To this end, we adopt the following standard assumptions in online convex optimization 
\cite{hazan2016introduction,agarwal2019online,zhang2018adaptive,gradu2020adaptive,zhao2021non,cao2018online,yi2020distributed}:

\begin{assumption}[Convex and Compact Bounded Domains]\label{assumption:bounded_set}
    The time-varying domain sets $\calX_t$, $t\in\TimeSeq$, are convex and compact; also, they are contained in a {bounded} set $\mathcal{X}$ contains the zero point and has diameter $D$; \ie $\mathbf{0} \in \calX$, and $\|\mathbf{x}-\mathbf{y}\|\; \leq D$ for all $\mathbf{x} \in \mathcal{X}, \mathbf{y} \in \mathcal{X}$.\footnote{An example of a bounded set $\calX$ containing all $\calX_t$, $t\in\TimeSeq$ is the $\calX=\calX_1\cup\cdots\cup\calX_T$. Then, $\calX$'s  diameter $D$ is finite since all $\calX_t$, $t\in\TimeSeq$, are compact.}
\end{assumption}

\Cref{assumption:bounded_set} considers time-varying domains $\calX_t$, $t\in\TimeSeq$, in contrast to the standard \OCO, which considers a time-\underline{in}variant domain $\calX$, \ie $\calX_1=\ldots=\calX_T=\calX$.

\begin{assumption}[Convex Loss]\label{assumption:convexity}
    The loss function $f_t: \calX \rightarrow \mathbb{R}$ is convex in $\myx \in \calX$ for all $t\in\TimeSeq$.\footnote{The assumption can be relaxed such that the loss function $f_t: \calX_t \rightarrow \mathbb{R}$ is convex in $\myx \in \calX_t$.}
\end{assumption}

\begin{assumption}[Bounded Gradient]\label{assumption:gradient}
	The gradient norm of ${f}_t$ is at most $G$, where $G$ is a given non-negative number; \ie $\left\|\nabla {f}_t(\mathbf{x})\right\| \leq G$  for all $\mathbf{x} \in \mathcal{X}$ and $t \in \TimeSeq$.\footnote{The assumption can be relaxed such that the gradient $\left\|\nabla {f}_t(\mathbf{x})\right\| \leq G$  for all $\mathbf{x} \in \calX_1 \cup \cdots \cup \calX_T$.}
\end{assumption}

\begin{algorithm}[t]
	\caption{\mbox{\hspace{-.05mm}Safe Online Gradient Descent (\SafeOGD)}.}
	\begin{algorithmic}[1]
		\REQUIRE Time horizon $T$; step size $\eta$.
		\ENSURE Decision $\mathbf{x}_t$ at each time step $t=1,\ldots,T$.
		\medskip
		\STATE Initialize $\mathbf{x}_1 \in \calX_1$; 
		\FOR {each time step $t = 1, \dots, T$}
		\STATE Suffer a loss $f_t(\mathbf{x}_t)$;
		\STATE Obtain gradient $\nabla f_t(\mathbf{x}_t)$;
		\STATE Obtain domain set $\calX_{t+1}$;
		\STATE Update $\myx_{t+1}^\prime=\myx_t- \eta \nabla f_t(\myx_t)$;
		\STATE Project $\myx_{t+1} = \Pi_{\calX_{t+1}}(\myx_{t+1}^\prime)$;
		\ENDFOR
	\end{algorithmic}\label{alg:SafeOGD}
\end{algorithm}

We present \SafeOGD (\Cref{alg:SafeOGD}), the first algorithm with bounded dynamic regret for \SafeOCO (\Cref{prob:safe_OCO}). 
\SafeOGD first takes as input the time horizon $T$ and a constant step size $\eta$, and initializes $\myx_1 \in \calX_1$ (line 1). At each time step $t$, \SafeOGD chooses a decision $\myx_t$, then suffers a loss $f_t(\myx_t)$ and evaluates the gradient $\nabla f_t(\mathbf{x}_t)$ (lines 3-4). The new domain set $\calX_{t+1}$ is then revealed and the algorithm performs the update step $\myx_{t+1}^\prime=\myx_t- \eta \nabla f_t(\myx_t)$ and projection step $\myx_{t+1} = \Pi_{\calX_{t+1}}(\myx_{t+1}^\prime)$ to compute the new decision $\myx_{t+1}$ (lines 5-7).

\begin{remark}[\SafeOGD \vs~\OGD]
    \SafeOGD generalizes the seminal \OGD to handle time-varying domain sets. Compared to \OGD where the domain set is time-invariant, \SafeOGD needs to obtain a changing domain set $\calX_{t+1}$ at every iteration (line 5) and project the intermediate decision $\myx_{t+1}^\prime$ into $\calX_{t+1}$ in the project step to satisfy the time-varying constraints (line 7). The challenge is that $\calX_{t+1}$ may be disjoint from the previous domain set $\calX_t$. The comparison between \OGD and \SafeOGD is illustrated in \Cref{fig:OGDvsSafeOGD}. 
\end{remark}

\begin{figure}[t]
    \centering
    \includegraphics[width=0.49\textwidth]{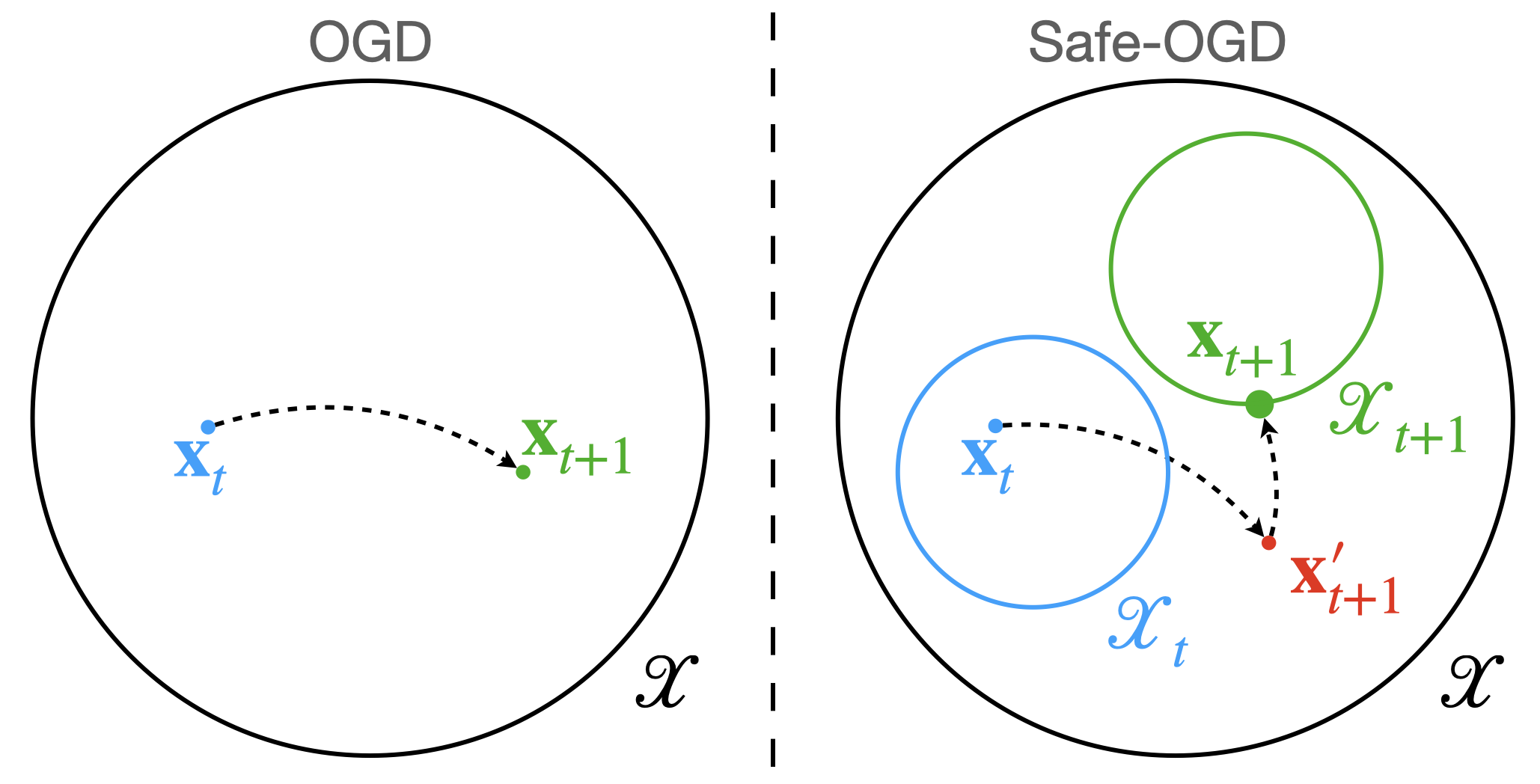}
    \caption{\textbf{Illustration of differences between \OGD and \SafeOGD}. \OGD updates the decision $\myx_t$ to $\myx_{t+1} \in \calX$.   \SafeOGD instead finds $\myx_{t+1} \in \calX_{t+1}$: it first updates $\myx_{t} \in \calX_{t}$ to $\myx_{t+1}^\prime$ (line 6 in \Cref{alg:SafeOGD}), and then projects $\myx_{t+1}^\prime$ to $\myx_{t+1} \in \calX_{t+1}$ (line~7).
    }
    \label{fig:OGDvsSafeOGD}
\end{figure}

We present dynamic regret bounds for \SafeOGD against any comparator sequence (\Cref{theorem:SafeOGD}), also demonstrating that the regret bounds reduce to those in standard OCO setting when the domain sets are time-\underline{in}variant (\Cref{remark:bound_reduction}).
We use the notation:
\begin{itemize}
    \item $\bar{\myx}_{t+1} \triangleq \Pi_{\calX_{t}}(\myx_{t+1}^\prime)$ is the decision would have been chosen at time step $t+1$ if $\calX_t = \calX_{t+1}$;

    \item $\zeta_t \triangleq \left\|\bar{\myx}_{t+1} - {\myx}_{t+1} \right\| = \left\|\Pi_{\calX_{t}}(\myx_{t+1}^\prime) - \Pi_{\calX_{t+1}}(\myx_{t+1}^\prime) \right\|$ is the distance between $\bar{\myx}_{t+1}$ and ${\myx}_{t+1}$, which are the projection of ${\myx}_{t+1}^\prime$ onto sets $\calX_t$ and $\calX_{t+1}$, respectively. $\zeta_t$ becomes $0$ when $\calX_t=\calX_{t+1}$;

    \item $S_T \triangleq \SumOneT \zeta_t$ is the cumulative variation of decisions due to time-varying domain sets. $S_T$ becomes $0$ when domain sets are time-\underline{in}variant; 

    \item { $C_T \triangleq \sum_{t=2}^{T} \| \myv_{t-1} - \myv_{t} \|$ is the path length of the sequence of comparators.}
\end{itemize}

We have the following regret bound of \SafeOGD.
\begin{theorem}[Dynamic Regret Bound of \SafeOGD]\label{theorem:SafeOGD}
Consider the Safe OCO problem. \SafeOGD achieves against any sequence of comparators $(\mathbf{v}_1, \dots, \mathbf{v}_T) \in \calX_1 \times \cdots \times \calX_T$ 
\begin{equation}
    \DyReg \leq \frac{\eta T G^2}{2} + \frac{7 D^2}{4 \eta} + \frac{D C_T}{\eta} + \frac{D S_T}{\eta}.
    \label{eq:theorem_SafeOGD_bound_init_app}
\end{equation}
Specifically, for $\eta=\calO\left(\frac{1}{\sqrt{T}}\right)$, 
\begin{equation}
    \DyReg \leq \calO\left(\sqrt{T}\left(1+C_{T}+S_{T}\right)\right).
   	\label{eq:theorem_SafeOGD_bound_app}
\end{equation}
\end{theorem}

The dependency on $C_T$ results from the sequence of comparators being time-varying.
Specifically, \cite{zhang2018adaptive} proved that any optimal dynamic regret bound  for \OCO is $\Omega\left(\sqrt{T(1+C_T)}\right)$, and thus the bound necessarily depends on $C_T$ in the worst case. 

The dependency on $S_T$ results from the domain sets being time-varying. $S_T$ is zero when the domain sets are time-\underline{in}variant (\Cref{remark:bound_reduction});  thus, $S_T$ can be sublinear in decision-making applications where any two consecutive safe sets differ a little (\eg in high-frequency control applications where the control input is updated every a few tenths of milliseconds, then the collision-free space may change only a little between consecutive time steps).

When the domain sets time-\underline{in}variant, the regret bounds in \cref{eq:theorem_SafeOGD_bound_app} reduce to the results in the standard \OCO setting, per the following remark.

\begin{remark}[Regret Bounds in the Time-\underline{In}variant Domain Case]\label{remark:bound_reduction}
    When the domain sets are time-\underline{in}variant, \ie $\calX_1 = \dots = \calX_T$, we have $S_T = 0$ by definition. Hence, the dynamic regret bounds in \cref{eq:theorem_SafeOGD_bound_app} reduce to $\calO\left(\sqrt{T}\left(1+C_{T}\right)\right)$, \ie they become equal to the dynamic regret bounds of \OGD in the standard \OCO setting  \cite{zinkevich2003online}.
\end{remark}



\end{document}